\documentclass[aps, pra, twocolumn, superscriptaddress, amsmath, amssymb]{revtex4-2}

\usepackage{graphicx}
\usepackage{dcolumn}
\usepackage{bm}
\usepackage{mathptmx} 
\newcommand*{\hham}{\mathcal{H}}
\newcommand*{\lham}{\mathcal{L}}
\usepackage{color}
\usepackage[dvipsnames]{xcolor}
\usepackage[hidelinks,colorlinks=true, linkcolor=WildStrawberry, urlcolor=WildStrawberry, citecolor=Emerald, linktocpage=true, breaklinks=true, bookmarksopen=true]{hyperref}
\usepackage{multirow}
\usepackage[USenglish]{babel}
\begin{document}

	\title[Multipolar and nonlocal effects in plasmon-mediated entanglement generation]
	{Multipolar and nonlocal effects in plasmon-mediated entanglement generation} 
	
	\author{Luke C. Ugwuoke}
	\email{ lcugwuoke@gmail.com }
	\affiliation{
		Department of Physics, Stellenbosch University, 
		Private Bag X1, Matieland 7602, South Africa.}
	
	\author{Tjaart P. J. Kr\"{u}ger}
	\affiliation{ 
		Department of Physics, University of Pretoria, 
		Private Bag X20, Hatfield 0028, South Africa.}
	\affiliation{National Institute for Theoretical and Computational Sciences (NITheCS), South Africa.}
	\author{Mark S. Tame}
	\affiliation{
		Department of Physics, Stellenbosch University, 
		Private Bag X1, Matieland 7602, South Africa.}
	\date{\today}
	\hyphenpenalty = 1000
	\begin{abstract}
		The generation of quantum entanglement is important for a wide range of quantum
technology protocols. In nanophotonics, a promising platform for quantum technologies, entanglement generation via plasmon-mediated coupling in quantum dot qubits is often modeled within the dipole limit, where only dipolar plasmons of the mediating nanoparticle are considered, and the local response approximation, where nonlocal corrections are ignored. However, multipolar effects manifest strongly at coupling distances less than the nanoparticle size, while nonlocal optical effects stem from a size-induced dielectric response. We investigated these two important effects in the generation of two-qubit entanglement mediated by plasmonic coupling. A cavity quantum electrodynamic approach is employed, where the induced plasmonic effects lead to modified transition rates in the dynamics of the coupled quantum dot qubits. 
We find that multipolar modes and size-dependent damping lead to entanglement decay at small coupling distances and limit mediated entanglement with certain particle sizes. We discuss potential implications of multipolar modes in entanglement-based quantum sensing. 
	\end{abstract}
	
	\maketitle 
	
	\section{Introduction}\label{section1}
	The electronic or spin states of certain semiconducting materials, having small dipole moments, and possessing two possible energy levels corresponding to the energy of an excited state (spin up) and a ground state (spin down), can be represented as a two-level system referred to as a \emph{quantum dot qubit} \cite{Eriksson2013}.
	When two or more single quantum dot qubits interact, they can exist in a strongly correlated and shared quantum state, an \emph{entangled state} \cite{Dur2000,Tanas2004,Sinaysky2008}. The degree of entanglement between two interacting qubits---bipartite entanglement---can be measured using Wootters' \emph{concurrence}---a dimensionless quantity that depends on the density matrix of the two-qubit system \cite{Wootters1998}. Entangled states have been used for a variety of tasks in quantum technology, including in securing quantum communication channels \cite{Schimp2021,Forbes2024}, improving image quality at low light intensities \cite{Venegas2010,Moodley2024}, and enhancing sensing resolution \cite{Lee16,Cres12}. 
	
	The entangling of two quantum dot qubits is largely dependent on their dipole--dipole coupling strength \cite{Tanas2004}, which has an inverse law dependence on the inter-qubit separation. In plasmon-mediated entanglement, the qubits interact via a mediated coupling strength due to a metal nanoparticle (MNP) \cite{Hou2014, He2012, Nerk2015, Dumit2017}, or a plasmon waveguide \cite{Martin2-2011, Martin2011, Li2019, Li2021, Fang2025, Du2026, Qiu2024, Hensen2018, Ugwuoke2025}, placed between the quantum dots (QDs). 
	Within the dipole approximation, $d \ge 2r$, where $d$ is the center-to-center distance between each QD and the MNP and $r$ is the MNP radius, previous studies have shown that mediated interactions can be used to generate stationary \cite{Hou2014} and transient entanglement \cite{He2012,Otten2015}. The process involves coherent driving of the MNP-QD system at the localized surface plasmon resonance (LSPR) wavelength $\lambda = \lambda_{0}$ of the dipole plasmon, while taking into account the mediated interactions---mediated coupling and cross-decay \cite{Hou2014, He2012, Nerk2015, Dumit2017, Otten2015, Otten2019}. 
	
	As was previously shown in Refs. \cite{Zhao2015,Niko2016,Hakami2016}, multipolar effects can become significant in MNP-QD interactions beyond the dipole limit. Multipolar modes include the quadrupole mode and higher-order modes, which cannot couple directly to a uniform external field due to their lack of a dipole moment; rather, they couple to the dipole mode, leading to their indirect excitation \cite{Zhao2015,Ugwuoke2022}. Hence, they are usually referred to as \emph{dark modes} \cite{Ugwuoke2022,Gomez2013,Hakala2017,Yang2017}. 
	
	On the other hand, the optical response of a MNP to an applied electric field has been shown to deviate considerably from size-independent dielectric functions \cite{Vial2005,Barch2014}, referred to as the \emph{local-response approximation} \cite{Raza2014,Raza2015,Hapu2017}, when particle sizes and separation distances approach the mean free path of the conduction electrons of the metal
	\cite{Fuch1981,Raza2014,Raza2015,Hapu2017, Raza2013-1,Moradi2015,Ye2025}. The deviation from the local response is due to size- and shape-dependent effects that emanate from convection, diffusion, and scattering of the surface conduction electrons of small MNPs, giving rise to a blueshift and linewidth broadening of the LSPR as well as a reduction in extinction efficiencies \cite{Raza2014,Raza2015,Raza2013-1}. 
	This has led to several implications for applied plasmonics, such as loss of coherence in plasmon lasing \cite{Malin2019}, suppression of plasmon-enhanced fluorescence \cite{Niko2016,Yong2020}, limitation of field confinement in plasmon waveguiding \cite{Raza2013-2}, and modifications of the optical response of MNP dimers \cite{Raza2012,Teper2013} and MNP-QD hybrid systems \cite{Hapu2017}. Tunable nonlocal effects have also been reported in graphene plasmonics \cite{Lund2017}.
	
	In this work, we investigate the effects of the multipolar and nonlocal optical responses of a MNP on qubit--qubit entanglement generation mediated by plasmonic coupling. This is done within the framework of cavity quantum electrodynamics, using an effective approach where the induced plasmonic effects result in modified detuning, coupling, driving, and dissipation terms in the dynamics of two plasmonically coupled QD qubits. The multipolar response allows us to conduct a comprehensive distance-dependent study, while the nonlocal response enables a size-dependent study involving small uncoated MNPs, unlike in previous work \cite{Hakami2016}. Although we restrict the study to the Rayleigh regime---where the particle size and coupling distance are small compared to the excitation wavelength---radiation damping due to retardation of the induced electric field on the QD by the MNP is taken into account phenomenologically, as given in Refs.~\cite{Waks2010,Ugwuoke2024}.
	
	This paper is organized as follows. The geometry and parameters of the model, including the QD and MNP parameters, multipolar and nonlocal terms, as well as the model Hamiltonian and Liouvillian, are introduced in Sec.~\ref{section2}. This is followed by an effective description of the two-qubit dynamics in Sec.~\ref{section3}. Entanglement generation via the mediated coupling and cross-decay rates is discussed in
    Sec.~\ref{section4}, based on nonlocal, dipole and multipole analyses of the concurrence. Quantum sensing based on the generated entanglement is examined in Sec.~\ref{section5} using the quantum Fisher information, and Sec.~\ref{section6} concludes the paper. 
	
	\section{Model}\label{section2}
	\begin{figure} 
		\centering 
		\includegraphics[width=0.45\textwidth]{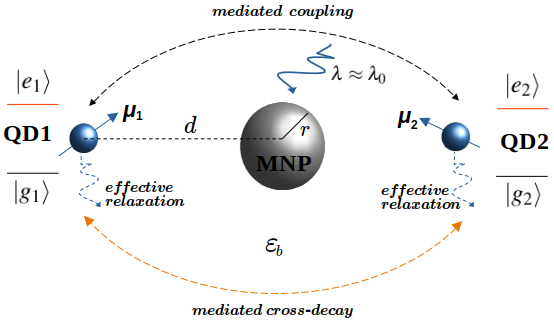}
		\caption{(Color online) Model geometry of the coupled system showing two QDs, QD1 and QD2, each with ground and excited states $|g_{i}\rangle$ and $|e_{i}\rangle~ (i = 1, 2)$ and
			dipole moments $\bm{\mu}_{1}$ and $\bm{\mu}_{2}$, respectively, coupled to a MNP of radius $r$, and interacting with one another via plasmon-mediated interactions at an inter-qubit distance of $2d$, and experiencing effective relaxation rates. The QD-MNP-QD system is driven at the dipolar LSPR wavelength $\lambda \approx \lambda_{0}$, of the MNP, in host medium of dielectric constant, $\varepsilon_{b}$.
		}\label{f1}
	\end{figure}
	
Fig. ~\ref{f1} shows the model geometry of the MNP-coupled QD qubits. Each QD is at a center-to-center distance $d = r_{0} +s + r$ from the MNP, where $r_{0}$ is the radius of each QD, $r$ is the MNP radius, and $s$ is the surface-to-surface distance between each QD and the MNP. 
Each QD has a dipole moment given by $\bm{\mu}_{i} = \mu\hat{e}_{i}$, $i = 1, 2$, with magnitude $\mu = er_{0}$, $e$ the electronic charge, and $\hat{e}_{i}$ a unit vector indicating the dipole orientation. We denote the dipole moment of the MNP by $\bm{\chi} = \chi\hat{e}_{m}$. 
We only consider QDs whose sizes are small in comparison to the MNP, i.e., $r_{0}<<r$, allowing us to treat the QDs as point dipoles modeled as two-level quantum emitters, as have been done in previous works \cite{Zhao2015,He2012}.
The QD-MNP-QD system is driven by a time-varying and uniform (quasi-static) field, $\bm{E} \approx E_{0}\cos(\omega t)\hat{e}$, where $E_{0}$ is the amplitude of the field, $\omega$ is the driving frequency, and $\hat{e}$ is the electric field polarization vector. We assume that the unit vectors, $\hat{e}, \hat{e}_{i}$, and $\hat{e}_{m}$ are parallel to the interaction axis of the QD-MNP-QD system. As shown previously, this configuration leads to optimal coupling \cite{Zhao2015,Ugwuoke2024} and a high MNP-QD cooperativity parameter \cite{Waks2010}. 

We denote $\omega^{L}_{l}, g^{L}_{l}$, and $\gamma^{L}_{l}$ as the LSPR frequency, the QD-MNP coupling rate, and the non-radiative plasmon damping rate of the MNP, respectively, due to contributions from both bright ($l = 1$) and dark ($l\ge 2$) modes, with $l$ being the angular momentum number, within the local-response approximation. 
The plasmon resonances of multipolar modes, $\omega_{l \ge 2}$, are blueshifted from the LSPR of the dipole mode, $\omega_{l=1}$, i.e., they are higher-energy modes, as shown in Refs.~\cite{Zhao2015,Raza2015,Ugwuoke2022}. Their contribution to the mediated coupling is through the dipole--multipole coupling strength, $g^{L}_{l\ge 2}$, which is a result of their interaction with the QD dipole, as given in Ref.~\cite{Zhao2015}. 

\subsection{Multipolar parameters and nonlocal corrections}\label{subsection2.1}
Following the results derived in Refs.~\cite{Raza2014,Raza2015,Hapu2017,Malin2019}, the local parameters, in addition to the magnitude of the MNP dipole moment, $\chi^{L}$, acquire size-dependent correction terms due to nonlocal effects, as follows: 
\begin{subequations}
\begin{align}
\omega_{l} & \approx \omega^{L}_{l} + \sqrt{l(l+1)}\frac{\beta}{2r}, \label{e1a} \\
g_{l}      & \approx g^{L}_{l}\sqrt{\Re(1+\Delta_{l})}, \label{e1b} \\
\gamma^{nr}_{l} & \approx \gamma^{L}_{l} + \frac{l}{2}\sqrt{\frac{l+1}{2l+1}}\frac{D\omega_{p}}{\beta r}, \label{e1c} \\
\chi & \approx \chi^{L}\sqrt{\Re(1+\Delta_{l=1})}, \label{e1d} 
\end{align}
\end{subequations}
where the local parameters derived in Refs.  \cite{Zhao2015,Ugwuoke2024} are: 
\begin{subequations}
\begin{align}
\omega^{L}_{l} &  =  \sqrt{ \frac{l\omega^{2}_{p} }{l\varepsilon_{\infty} + (l+1)\varepsilon_{b}} - \gamma^{2}_{p}  }, \label{e2a} \\
g^{L}_{l} &  = \frac{\mu(l+1)}{d^{l+2}}\sqrt{ \frac{(2l+1)\eta_{l}r^{2l+1} }{4\pi\varepsilon_{0}\hbar l } }, \label{e2b} \\
\gamma^{L}_{l} &  = \gamma_{p}\left[1  + \Big( \frac{\gamma_{p}}{\omega^{L}_{l}} \Big)^{2} \right], \label{e2c} \\
\chi^{L} &  = \epsilon_{b}\sqrt{12\pi\epsilon_{0} \eta_{l=1}r^{3}\hbar} \label{e2d}. 
\end{align}
\end{subequations}
In Eqs.~\eqref{e1a} and \eqref{e1c}, $\beta = \sqrt{3/5}v_{F}$ is the convective term proportional to the pressure acting on the electron density and $D = 4\gamma_{p}v^{2}_{F}/15(\omega^{2} + \gamma_{p}^{2})$ is the electron diffusion constant due to surface effects, in the limit $\omega >>\gamma_{p}$, of the dielectric response of the metal. Furthermore, $v_{F}$ is the Fermi velocity of the metal, 
$\omega_{p}$  and $\gamma_{p}$ are, respectively, the bulk frequency and damping rate of the free electrons of the metal, and dielectric effects have been ignored in the nonlocal terms \cite{Raza2015}. 
In Eqs.~\eqref{e1b} and \eqref{e1d}, the parameter $\Delta_{l}$ is the nonlocal 
correction, given by \cite{Raza2014,Raza2015} 
\begin{equation}\label{e3}
\Delta_{l} = l(l+1)\frac{\varepsilon(\omega) - \varepsilon_{\infty}(\omega) }{\varepsilon_{\infty}(\omega)}\frac{j_{l}(k_{L}r)}{k_{L}r j_{l}^{\prime}(k_{L}r)},
\end{equation}
where $j_{l}(k_{L}r)$ is the spherical Bessel function of the first kind and $j_{l}^{\prime}(k_{L}r)$ is the first derivative of this function with respect to the size parameter $k_{L}r$, with $k_{L} = \sqrt{\omega(\omega+i\gamma_{p}) \varepsilon(\omega)/\varepsilon_{\infty}(\omega)[\beta^{2} + D(\gamma_{p}-i\omega)] }$ being the wavevector of the longitudinal electric field. 
The local dielectric function, 
$\varepsilon(\omega) = \varepsilon_{\infty}(\omega) - \omega^{2}_{p}/\omega(\omega+i\gamma_{p})$, represents a combination of the response of the bulk plasmons of the metal due to the polarization of the positive ion core and bound electrons, represented by the high-frequency dielectric function $\varepsilon_{\infty}(\omega)$, and the polarization of the free electrons, represented by the Drude term $\omega^{2}_{p}/\omega(\omega+i\gamma_{p})$ \cite{Vial2005,Barch2014}. Here, we will use a constant value for $\varepsilon_{\infty}(\omega)$, i.e., $\varepsilon_{\infty}(\omega) \approx \epsilon_{\infty}$, 
with the real part of $\varepsilon_{\infty}(\omega)$ ranging from $3$ to $5$ at optical frequencies, for example, in silver \cite{Barch2014}. 
Note that in Eq. \eqref{e1b}, we have replaced the dipolar coupling strength $g_{l=1}$ and nonlocal correction term $\Delta_{l=1}$ in Eq. (47a) of Ref.~\cite{Hapu2017} with the multipolar coupling strength $g_{l}$ and nonlocal correction term $\Delta_{l}$ from 
Ref.~\cite{Zhao2015} and Ref.~\cite{Raza2015}, respectively. Though this is a phenomenological treatment, it allows us to discuss multipolar contributions. The approximation is plausible since the local term is recovered in the limit $\Delta_{l} \rightarrow 0$. 
In Eqs. \eqref{e2a} and \eqref{e2d}, $\varepsilon_{b}$ is the dielectric constant of the host medium. In Eqs. \eqref{e2c} and \eqref{e2d}, $\varepsilon_{0}$ is the free-space permittivity, $\hbar$ is the Dirac constant, and $\eta_{l} = (1/2\omega_{l})(l\omega_{p}/[l\varepsilon_{\infty} + (l+1)\varepsilon_{b} ])^{2}$. 
We denote the total plasmon damping by $\gamma_{l}$, so that
$\gamma_{l} = \gamma^{nr}_{l} + \gamma^{r}_{l}\delta_{1l}$ accounts for both the non-radiative and radiative components of the plasmon decay. The decay rate, $\gamma^{r}_{1} = \chi^{2}\sqrt{\varepsilon_{b}}\omega^{3}_{l=1}/3\pi\varepsilon_{0}\hbar c^{3}$, where $c$ is the speed of light in free space, is due to dipole radiation \cite{Waks2010,Ugwuoke2024}. 

In the following, we will use a cavity quantum electrodynamic framework \cite{Hou2014,Waks2010} to investigate the role of dark modes and nonlocal effects in plasmon-mediated entanglement while restricting the model to the weak-excitation regime, $\Omega << \gamma_{1}$, with $\Omega$ being the excitation rate of the MNP and $\gamma_{1}$ the decay rate of the dipole plasmon (corresponding to the mode order $l = 1$). This ensures the validity regime of the effective approach which we employ. As described in Ref. \cite{Zhao2015}, this approach enables a treatment of the multimode MNP-QD dynamics in a single-mode picture.

	\subsection{Model Hamiltonian and Liouvillian}\label{subsection2.2}
Based on previous works \cite{Zhao2015,Ugwuoke2024}, the following Hamiltonian is obtained for the QD-MNP-QD system in a frame rotating with the driving frequency $\omega$ of the applied field: 
\begin{eqnarray}\label{e4}
	\hham &=&  \hbar\sum_{l = 1}^{N}\left(\Delta\omega_{l} a_{l}^{\dagger}a_{l}-\Omega_{l}\delta_{1l}(a_{l}^{\dagger} + a_{l})\right) \nonumber \\
	& & + \hbar\sum_{i = 1}^{2} \Delta\omega_{i}\sigma^{\dagger}_{i}\sigma_{i} -\hbar \sum_{i = 1}^{2}\sum_{l = 1}^{N}g_{il}(\sigma^{\dagger}_{i}a_{l} + \sigma_{i}a_{l}^{\dagger}), 
\end{eqnarray}
where $N$ is the number of multipoles, $\Delta\omega_{l} = \omega_{l}-\omega$ is the detuning of the plasmon resonances, $\omega_{l}$, of the MNP, from the driving frequency, $a_{l}^{\dagger}(a_{l})$ is the creation (annihilation) operator of the MNP multipole plasmons, $\Delta\omega_{i} =\omega_{i}-\omega$ is the detuning of the transition frequency, $\omega_{i}$, of each QD, from the driving frequency, $\sigma_{i}^{\dagger} = |e_{i}\rangle \langle g_{i}|$ ($\sigma_{i} = |g_{i}\rangle \langle e_{i}|)$ is the raising (lowering) operator of each QD ($i = 1,2$), and $\Omega_{1} = \Omega = E_{0}\chi/2\hbar$ is the excitation rate of the MNP. 
The Liouvillian accounting for the spontaneous emission of each QD as well as the multipole plasmon damping is 
\begin{eqnarray}\label{e5}
	\lham[\rho] &=& \sum_{i = 1}^{2}\frac{\gamma_{i}}{2}(2\sigma_{i}\rho\sigma_{i}^{\dagger} - \{\sigma_{i}^{\dagger}\sigma_{i},\rho\}) \nonumber  \\ 
	& & + \sum_{l = 1}^{N}\frac{\gamma_{l}}{2}(2a_{l}\rho a_{l}^{\dagger} - \{a_{l}^{\dagger}a_{l},\rho\}),
\end{eqnarray}
where $\rho$ is the reduced density matrix of the system, $\gamma_{i}$ is the spontaneous emission rate of each QD, and we have modified the Liouvillian in Ref. \cite{Zhao2015} to account for the two QDs in our model. 

	\section{Effective description}\label{section3}
The equations of motion of the expectation values of the plasmon annihilation operator and the single-qubit lowering and population difference operators are obtained, respectively, using 
$\langle \dot{O} \rangle = tr(O\dot{\rho})$ and $\dot{\rho} = i[\rho, \mathcal{H}]/\hbar + \lham[\rho]$, as follows: 
\begin{subequations}
	\begin{align}
		\langle \dot{a}_{l} \rangle & = -(i\Delta\omega_{l} + \frac{1}{2}\gamma_{l})\langle a_{l} \rangle + i(\Omega_{1} +\sum_{i = 1}^{2}g_{il}\langle \sigma_{i} \rangle  ), 
		\label{e6a} \\
		\langle \dot{\sigma}_{i} \rangle & = -(i\Delta\omega_{i} + \frac{1}{2}\gamma_{i})\langle \sigma_{i}\rangle -i\sum_{l = 1}^{N}g_{il}\langle a_{l} \sigma^{z}_{i}\rangle
		\label{e6b}, \\
		\langle \dot{\sigma^{z}}_{i} \rangle & = -\gamma_{i}(\langle \sigma^{z}_{i} \rangle + 1  ) - 2i \sum_{l = 1}^{N}g_{il}(\langle a^{\dagger}_{l} \sigma_{i}\rangle - \langle a_{l} \sigma^{\dagger}_{i}\rangle  )
		\label{e6c}. 
	\end{align}
\end{subequations}
In order to arrive at an effective picture for the qubits, we need to employ two approximations. Firstly, we use a \emph{semiclassical approximation} \cite{Waks2010}, which allows the factoring of the joint expectation values of the MNP field operator and the single qubit operator, i.e., $\langle a_{l} \sigma^{z}_{i}\rangle \approx \langle a_{l}\rangle \langle \sigma^{z}_{i}\rangle$, and so on. Secondly, we solve for the stationary solution of Eq.~\eqref{e6a}, and use it to eliminate
$\langle a_{l}\rangle$ (and $\langle a^{\dagger}_{l}\rangle$) in the factored expectation values in Eqs.~\eqref{e6b} and \eqref{e6c}. This second approximation is known as \emph{adiabatic elimination} \cite{Zhao2015,Hou2014}. It allows the MNP dynamics to be treated as stationary with respect to the dynamics of the qubits. This is possible due to the fast plasmon dynamics ($\sim$ps to $\sim$fs timescales) compared to the dynamics of the qubits ($\sim$ns timescales).  
Based on these approximations, which are valid in the weak-excitation regime \cite{Waks2010,Hou2014}, beyond which they lead to under-prediction of the population of the Bell states \cite{Hou2014}, we obtain the following effective parameters of each of the two qubits: 
\begin{equation}
	\tilde{\Omega}_{i} = \sum_{l=1}^{N} g_{il}
	\frac{i \tilde{\Omega}_{l}\delta_{1l} }{\delta_{l}},~~
	\tilde{G}_{ij} = \sum_{l=1}^{N} \Delta\omega_{l}
	\frac{ g_{il}g_{jl} }{|\delta_{l}|^{2}},~~
	 \tilde{\Gamma}_{ij}  = \sum_{l=1}^{N} \gamma_{l}
	\frac{ g_{il}g_{jl} }{|\delta_{l}|^{2}}, \label{e7}\\
\end{equation}
\begin{equation}
	\Delta\tilde{\omega}_{i}  = \Delta\omega_{i} - \sum_{l=1}^{N} \Delta\omega_{l}
	\frac{ g_{il}^{2} }{|\delta_{l}|^{2}}, ~~
	\tilde{\gamma}_{i} = \gamma_{i} + \sum_{l=1}^{N} \gamma_{l}
	\frac{ g_{il}^{2} }{|\delta_{l}|^{2}}, \label{e8}
\end{equation}
where $\delta_{l} = i\Delta\omega_{l} +\frac{1}{2}\gamma_{l}$, and $\tilde{\Omega}_{i}, \tilde{G}_{ij}, \tilde{\Gamma}_{ij}, \Delta\tilde{\omega}_{i}$, 
$\tilde{\gamma}_{i}$ are the effective Rabi frequencies, coupling rates, cross-decay rates, detuning frequencies, and relaxation rates of each of the two qubits. 
Since $\gamma_{1}^r << \gamma_{1}^{nr}$, the cross-decay rates, $\tilde{\Gamma}_{ij}$, can be described as a decay process in which the qubits interact through non-radiative means. 
Together with $\tilde{G}_{ij}$, they constitute the \emph{plasmon-mediated interactions}, while $\Delta\tilde{\omega}_{i}$ (the \emph{exciton shift}) and 
$\tilde{\gamma}_{i}$ (the \emph{Purcell effect}) are plasmon-induced effects, and 
$\tilde{\Omega}_{i}$ are the plasmon-mediated Rabi frequencies. 
These parameters enable the construction of the effective two-qubit Hamiltonian and Liouvillian as: 
\begin{equation}
		\tilde{\mathcal{H}} = \hbar\sum_{i=1}^{2}\left(\Delta\tilde{\omega}_{i}\sigma_{i}^{\dagger}\sigma_{i} - (\tilde{\Omega}_{i}\sigma_{i}^{\dagger} + \tilde{\Omega}_{i}^{*}\sigma_{i} )  \right) 
- \hbar\sum_{i\neq j}^{2}\tilde{G}_{ij}\sigma_{i}^{\dagger}\sigma_{j}, \label{e9}
\end{equation}
\begin{equation}
		\tilde{\mathcal{L}_{ij}}[\rho]  = \frac{1}{2}\sum_{i,j=1}^{2} \tilde{\Gamma}_{ij}(2\sigma_{i}\rho\sigma_{j}^{\dagger} - \{\sigma_{i}^{\dagger}\sigma_{j},\rho\}), \label{e10}
\end{equation}
where in Eq.~\eqref{e10}, $i = j$ corresponds to spontaneous emission, and 
cross-decay otherwise. 
    \subsection{Effective parameters in the Dicke basis}\label{subsection3.1}
\begin{figure}
	\centering 
	\includegraphics[width=0.35\textwidth]{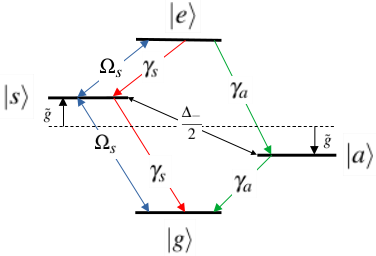}
	\caption{(Color Online) Energy level diagram showing the state transitions amongst the Dicke states and the rates involved.}
	\label{f2}
\end{figure}
 In the next subsection, we derive the equations of motion of the elements of the reduced density matrix of the two-qubit system in the computational basis: 
 $|1\rangle  = |g_{1}\rangle \otimes |g_{2}\rangle, |2\rangle  = |g_{1}\rangle \otimes |e_{2}\rangle, 
 |3\rangle  = |e_{1}\rangle \otimes |g_{2}\rangle,$ and $|4\rangle  = |e_{1}\rangle \otimes |e_{2}\rangle$. 
 However, it is useful for the discussion of bipartite entanglement to have an energy level diagram, as shown in Fig.~\ref{f2}, using the Dicke basis \cite{Tanas2004}: $|g\rangle  = |1\rangle, |s\rangle  = \frac{1}{\sqrt{2}}(|3 \rangle +|2 \rangle), 
 |a\rangle = \frac{1}{\sqrt{2}}(|3 \rangle - |2\rangle), |e\rangle  = |4\rangle$, with the effective parameters obtained as \cite{Hou2014}:
 $\Omega_{s} = \frac{1}{\sqrt{2}}(\tilde{\Omega}_{1} + \tilde{\Omega}_{2}), \Omega_{a} = \frac{1}{\sqrt{2}}(\tilde{\Omega}_{1} - \tilde{\Omega}_{2}), \Delta_{s} = \frac{\Delta_{+}}{2} + \tilde{g}, \Delta_{a} = \frac{\Delta_{+}}{2} - \tilde{g}, \Delta_{-} = \Delta\tilde{\omega}_{1} - \Delta\tilde{\omega}_{2}, \gamma_{s}  = (\tilde{\gamma}_{1} + \tilde{\gamma}_{2} + 2\tilde{\Gamma}_{12})/2, \gamma_{a}  = (\tilde{\gamma}_{1} + \tilde{\gamma}_{2} - 2\tilde{\Gamma}_{12})/2$, and $\Delta_{+} = \Delta\tilde{\omega}_{1} + \Delta\tilde{\omega}_{2}$. Here, $|g\rangle$ and $|e\rangle$ are the ground and excited states of the coupled qubits, and $|s\rangle$ and $|a\rangle$ are entangled states, referred to as \emph{symmetric and antisymmetric states}, respectively, and we have used $g_{1l} = g_{2l} \implies \tilde{G}_{12} = \tilde{G}_{21} = \tilde{g}$, since the MNP is centrally placed between the qubits, as shown in Fig.~\ref{f1}. This leads to $\tilde{\Omega}_{1} = \tilde{\Omega}_{2}  = \tilde{\Omega}$ and therefore $\Omega_{a} = 0$. Fig.~\ref{f2} is then drawn based on the Dicke Hamiltonian: $\Delta_{+}|e\rangle\langle e| + \Delta_{s} |s\rangle\langle s| + \Delta_{a} |a\rangle\langle a| + \frac{\Delta_{-} }{2}(|s\rangle\langle a| + |a\rangle\langle s| )-\Omega_{s}(|s\rangle\langle g| + |e\rangle\langle s|)-\Omega^{*}_{s}(|g\rangle\langle s| + |s\rangle\langle e|)$  and the rate equations for the populations in the Dicke basis (without driving): $\dot{\rho}_{ss}  \approx -\gamma_{s}(\rho_{ss} - \rho_{ee} ) -i\Delta_{-} (\rho_{as}  - \rho_{sa} )/2, ~\dot{\rho}_{aa}  \approx -\gamma_{a}(\rho_{aa} - \rho_{ee} ) + i\Delta_{-} (\rho_{as}  - \rho_{sa} )/2, ~\dot{\rho}_{gg}  \approx \gamma_{s}\rho_{ss} + \gamma_{a}\rho_{aa}, ~\dot{\rho}_{ee}  \approx -(\gamma_{s} + \gamma_{a})\rho_{ee}, \text{with} ~ \gamma_{a}  = \tilde{\gamma} - \tilde{\Gamma}_{12}, ~\text{and}~\gamma_{s}  = \tilde{\gamma} +  \tilde{\Gamma}_{12}$, where we have assumed that the qubits have the same spontaneous emission rate, leading to $\tilde{\gamma}_{1} = \tilde{\gamma}_{2} = \tilde{\gamma}$. 
	\subsection{Evolution of the reduced density matrix}\label{subsection3.2}
When driven by an external field via $\tilde{\Omega}_{i}$, the system evolves according to the Lindblad quantum master equation \cite{Tanas2004,Waks2010}: 
\begin{equation}\label{e11}
	\dfrac{\partial \rho}{\partial t} = \frac{i}{\hbar}[\rho, \tilde{\mathcal{H}}] + \tilde{\mathcal{L}_{ij}}[\rho].
\end{equation}
In the computational basis, the density matrix is given by
\begin{equation}\label{e12}
	\rho =
	\begin{bmatrix}
		\rho_{11} & \rho_{12} & \rho_{13} & \rho_{41}^* \\
		\rho_{12}^* & \rho_{22} & \rho_{23} & \rho_{42}^* \\
		\rho_{13}^* & \rho_{23}^* & \rho_{33} & \rho_{43}^* \\
		\rho_{41} & \rho_{42} & \rho_{43} & \rho_{44} \\
	\end{bmatrix},
\end{equation}
where we have used the complex conjugation property of the coherences: $\rho_{ij} = \langle i|\rho|j \rangle = \rho_{ji}^* = \langle j|\rho|i \rangle ^*$.  
 Solving Eq. \eqref{e11} for each density matrix element leads to the following equations: 
\begin{subequations} 
	\begin{align} 
		\dot{\rho}_{44} &= -2\tilde{\gamma} \rho_{44} + 2\Im\tilde{\Omega}^*(\rho_{43} + \rho_{42}), \label{e13a}\\ 
		\dot{\rho}_{33} &= -\tilde{\gamma}(\rho_{33} - \rho_{44}) - \tilde{\Gamma}_{12} \Re\rho_{23} + 2\tilde{G}_{12} \Im\rho_{23} \nonumber \\ &-2\Im(\rho_{43}\tilde{\Omega}^* + \rho_{13}\tilde{\Omega}), \label{e13c}\\
		\dot{\rho}_{22} &= -\tilde{\gamma}(\rho_{22} - \rho_{44}) - \tilde{\Gamma}_{12} \Re\rho_{23} - 2\tilde{G}_{12} \Im\rho_{23} \nonumber \\ &-2\Im(\rho_{42}\tilde{\Omega}^* + \rho_{12}\tilde{\Omega}), \label{e13d} \\
        \dot{\rho}_{11} &= -\tilde{\gamma}(\rho_{33} + \rho_{22}) + 2\tilde{\Gamma}_{12} \Re\rho_{23} + 2\Im\tilde{\Omega}(\rho_{12} + \rho_{13}), \label{e13b}\\ 
		\dot{\rho}_{41} &= -(i\tilde{\Delta}_{+} + \tilde{\gamma})\rho_{41} + i\tilde{\Omega}(\rho_{12}^* + \rho_{13}^* - \rho_{43} - \rho_{42}), \label{e13e}\\
		\dot{\rho}_{42} &= -(i\Delta\tilde{\omega}_{1} + 3\tilde{\gamma}/2)\rho_{42} + i\tilde{\Omega}(\rho_{22} - \rho_{44})\nonumber \\
		 &+ i(\rho_{23}\tilde{\Omega} - \rho_{41}\tilde{\Omega}^*) - (i\tilde{G}_{12} + \tilde{\gamma}/2)\rho_{43}, \label{e13g} \\
		\dot{\rho}_{43} &= -(i\Delta\tilde{\omega}_{2} + 3\tilde{\gamma}/2)\rho_{43} + i\tilde{\Omega}(\rho_{33} - \rho_{44})\nonumber \\
		& + i(\rho_{23}^*\tilde{\Omega} - \rho_{41}\tilde{\Omega}^*) - (i\tilde{G}_{12} + \tilde{\gamma}/2)\rho_{42}, \label{e13f} \\
		 \dot{\rho}_{12} &= (i\Delta\tilde{\omega}_{1} - \tilde{\gamma}/2)\rho_{12} - (i\tilde{G}_{12} + \tilde{\Gamma}_{12}/2)\rho_{13} \nonumber \\
		& + i\tilde{\Omega}^*(\rho_{22} - \rho_{11} + \rho_{23}^*) - i\tilde{\Omega}\rho_{41}^* + \tilde{\Gamma}_{12}\rho_{43}^* + \tilde{\gamma}\rho_{42}^*, \label{e13h}\\
		\dot{\rho}_{13} &= (i\Delta\tilde{\omega}_{2} - \tilde{\gamma}/2)\rho_{13} - (i\tilde{G}_{12} + \tilde{\Gamma}_{12}/2)\rho_{12} \nonumber \\
		& + i\tilde{\Omega}^*(\rho_{33} - \rho_{11} + \rho_{23}) - i\tilde{\Omega}\rho_{41}^* + \tilde{\gamma}\rho_{43}^* + \tilde{\Gamma}_{12}\rho_{42}^*, \label{e13i}\\
		\dot{\rho}_{23} &= -(i\tilde{\Delta}_{-} + \tilde{\gamma})\rho_{23} - (i\tilde{G}_{12} + \tilde{\gamma}/2)\rho_{33} + \tilde{\Gamma}_{12}\rho_{44} \nonumber \\
		& + (i\tilde{G}_{12} - \tilde{\gamma}/2)\rho_{22} - i\tilde{\Omega}(\rho_{43}^* - \rho_{12}) + i\tilde{\Omega}^*(\rho_{42} - \rho_{13}^*)\label{e13j}, 
	\end{align} 
\end{subequations}
where we have used the relation $\rho_{11} = 1 - (\rho_{22} + \rho_{33} + \rho_{44})$ to obtain Eq.~\eqref{e13b}.
	\section{Entanglement generation}\label{section4}
The degree of entanglement is calculated using the concurrence \cite{Wootters1998}:
\begin{equation}
	C = \text{max}\left(0, \lambda_{1} - \sum\limits_{k>1}\lambda_{k}  \right),
\end{equation}
with $\lambda_{k}$ being the eigenvalues in descending order of the matrix $\sqrt{ \sqrt{\rho}  \tilde{\rho} \sqrt{\rho}}$, with $\tilde{\rho} = (\sigma_{y} \otimes \sigma_{y})\rho^{*}(\sigma_{y} \otimes \sigma_{y})$, where $\sigma_{y}$ is the Pauli Y matrix and $\rho^{*}$ is the complex conjugate of $\rho$.

We consider two tunable cadmium-based QDs \cite{Deng2010}, each of radius $r_{0} = 0.8$ nm, relaxation rate $\gamma = 2\pi \times10^{8}$ rad/s, and transition frequencies $\omega_{1} = \omega_{pl} + \delta$ and $\omega_{2} = \omega_{pl} - \delta$, where $\omega_{pl} = \omega_{l=1}$ is the dipolar LSPR of the MNP. The QDs have been detuned anti-symmetrically via a small detuning frequency, $\delta = 10^{-5}\omega_{pl}$, to ensure 
population trapping in the slowly-decaying antisymmetric state $|a\rangle$ via a non-zero value of the detuning rate $\Delta_{-}$ in Fig.~\ref{f2}, as previously discussed in Refs.~\cite{Hou2014,He2012}. 
We have used $N = 10$, which was enough to achieve convergence in the multipole analysis,  i.e., multipoles of all orders from $l = 1$ to $l = 10$ were considered. In the dipole analysis, however, we only consider the contribution from the dipole mode, i.e., $l = 1$.

\subsection{Varying MNP-QD coupling distance}\label{subsection4.1}
The MNP considered is a silver nanoparticle (AgNP) of radius $r = 30$ nm in a host medium of dielectric constant $\varepsilon_{b} = 3.0$. 
This corresponds to $\lambda_{0} = 2\pi c/ \omega_{pl} \approx 481$ nm in the local-response approximation via Eq.~\eqref{e2a} (which can be obtained with the Drude parameters $\omega_{p} = 8.5472$ eV, $\varepsilon_{\infty} = 5$, and $\gamma_{p} = 0.018$ eV), and a size-induced blueshift with nonlocal corrections via Eq.~\eqref{e1a}, with $v_{F} = 1.39\times10^{6}$ m/s, leading to $\lambda_{0} \approx 478$ nm. The AgNP is driven resonantly by a laser field whose intensity is $I = 10$ W/cm$^{2}$, with the amplitude $E_{0} = (2I/cn\varepsilon_{0})^{1/2}$, corresponding to a weak-excitation regime $\Omega/\gamma_{1} \approx 0.02$.  
We start by investigating the dependence of the transient concurrence on the MNP-QD surface-to-surface distance, $s$, at a constant MNP radius, $r = 30$ nm, within the dipole and multipole analyses. 
The following initial states of the two QD qubits are considered: $\rho_{0} = |1\rangle\langle 1|$ and $\rho_{0} = |3\rangle\langle 3|$. 

 \begin{figure} 
	\centering 
	\includegraphics[width = 0.24\textwidth]{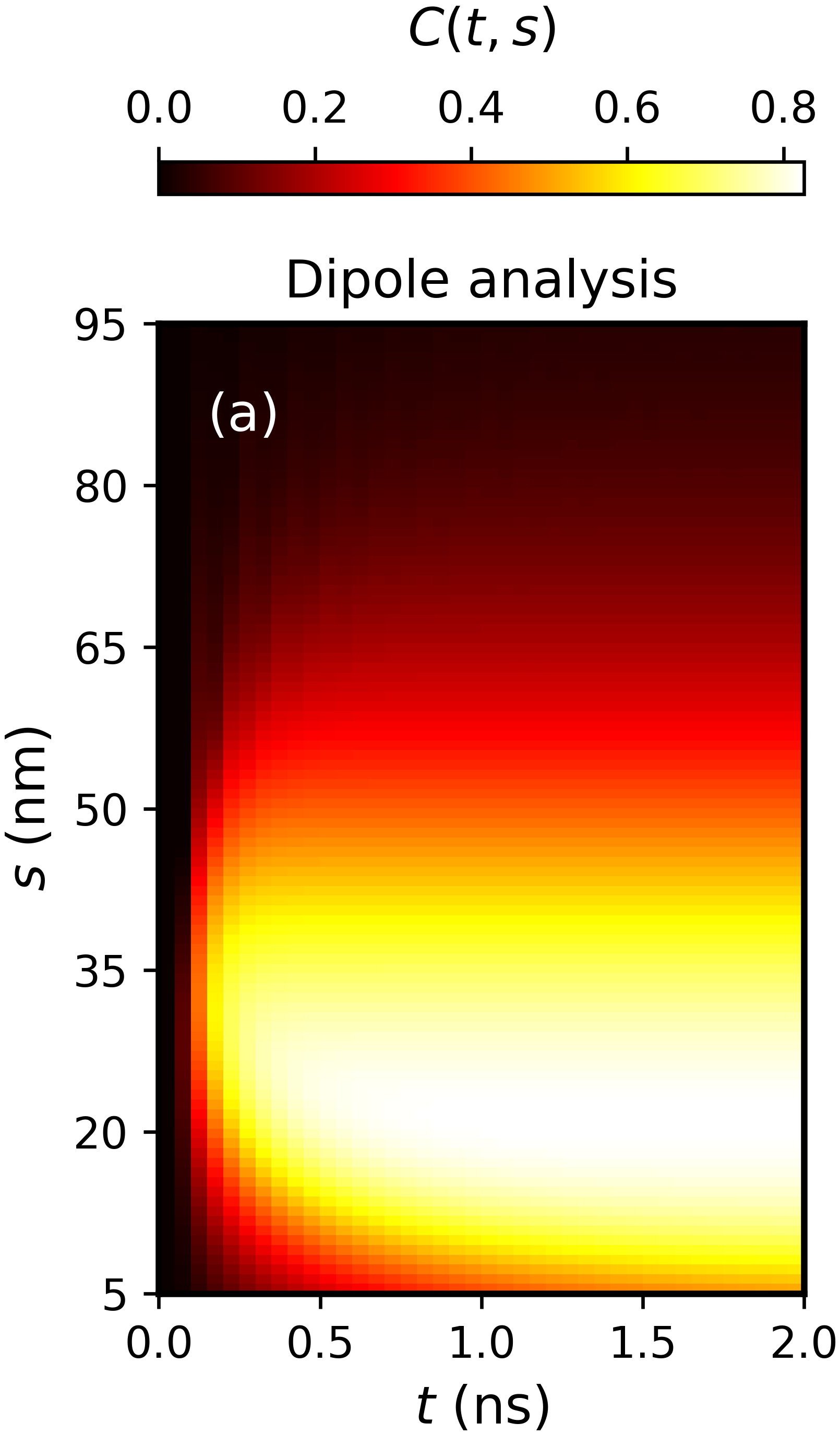}~~~~~~
    \includegraphics[width = 0.205\textwidth]{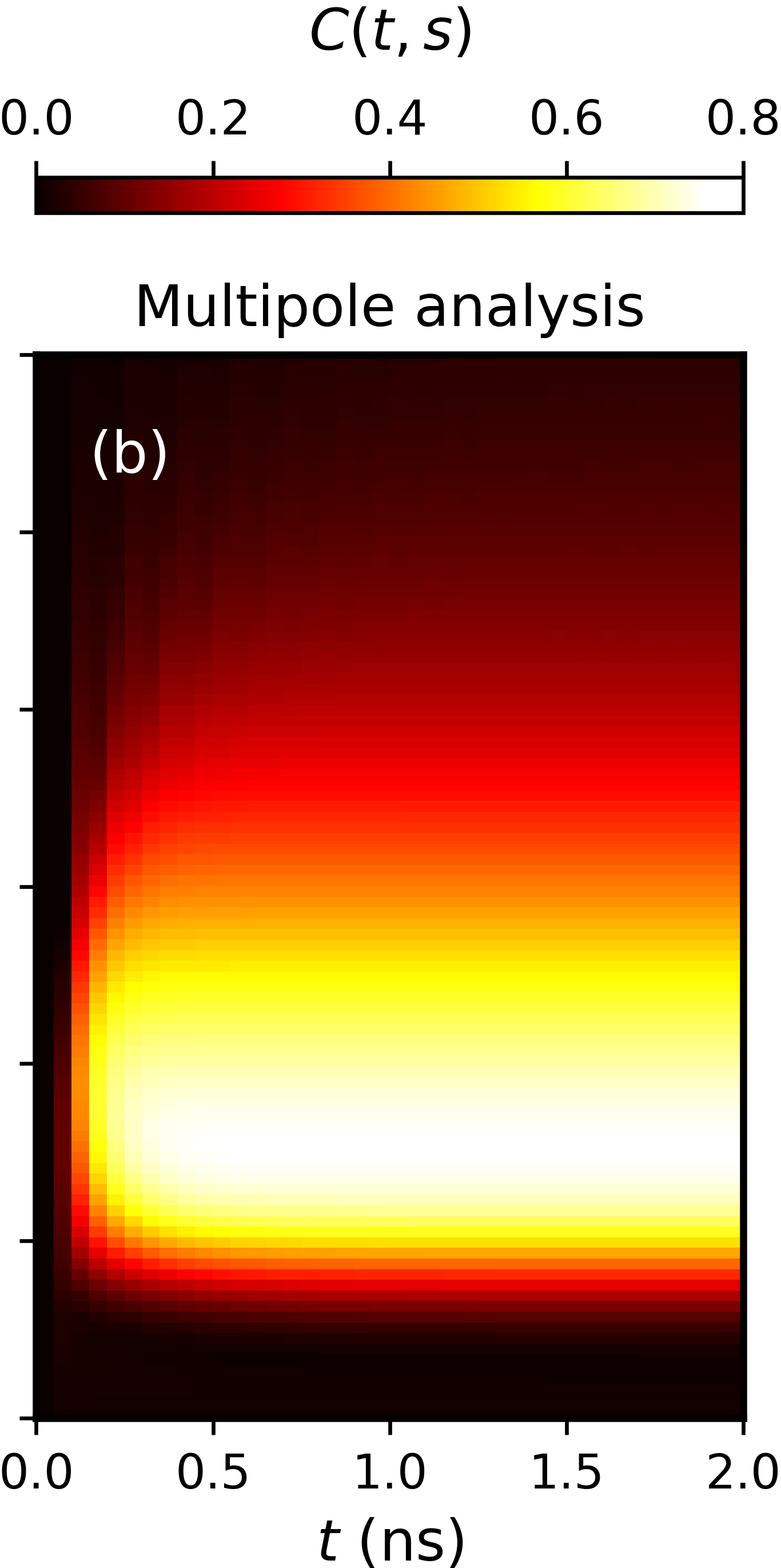}
	\caption{(Color online) Comparison of the dipole and multipole analyses of the transient concurrence of the two-qubit system for qubits prepared in the initial state $\rho_{0} = |1\rangle\langle 1|$ for different values of MNP-QD surface-to-surface distance $s$, and a AgNP of radius, $r = 30$ nm.
	}\label{f3}
\end{figure}
 \begin{figure}
	\centering 
	\includegraphics[width = 0.24\textwidth]{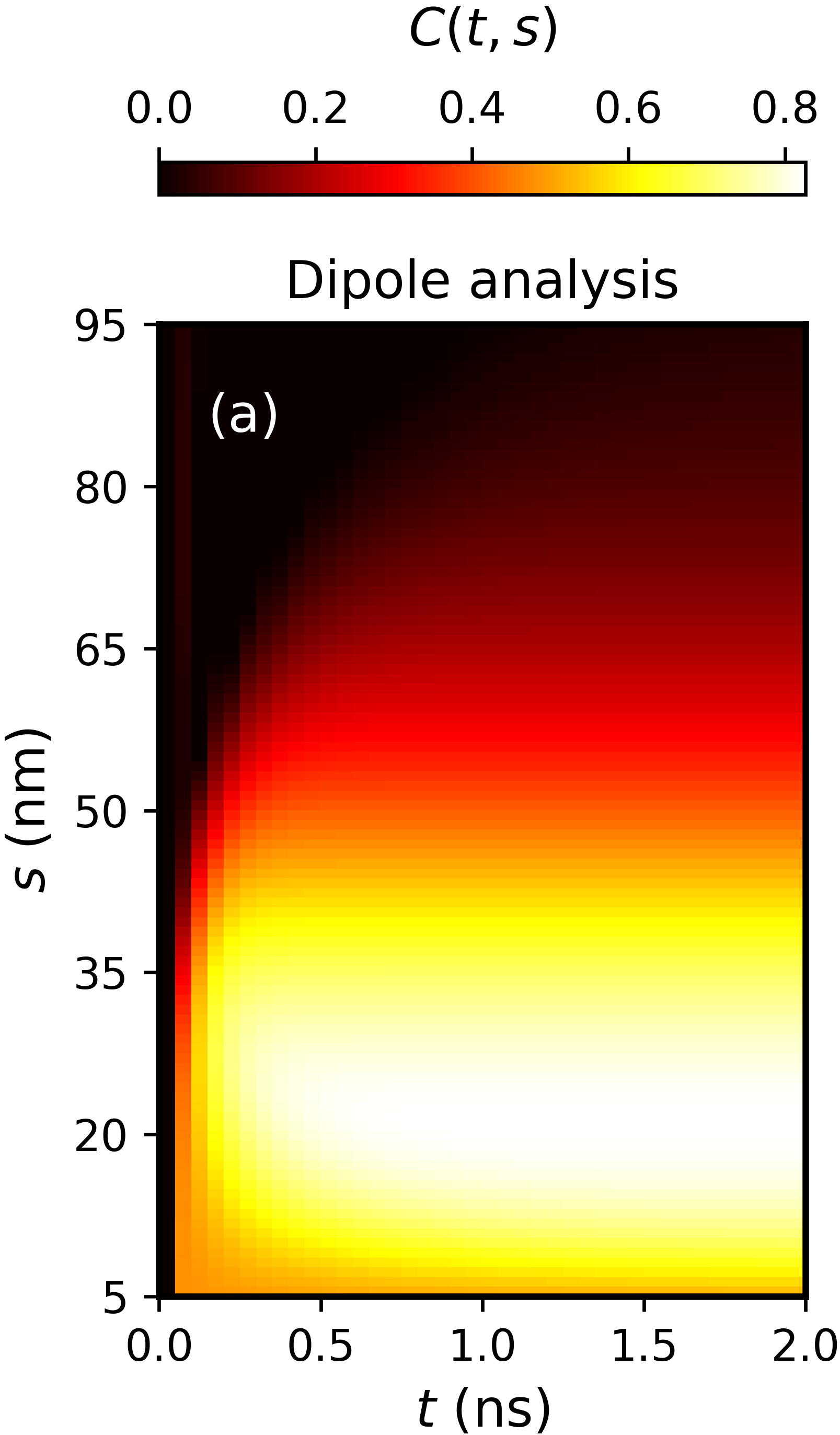}~~~~~~
    \includegraphics[width = 0.205\textwidth]{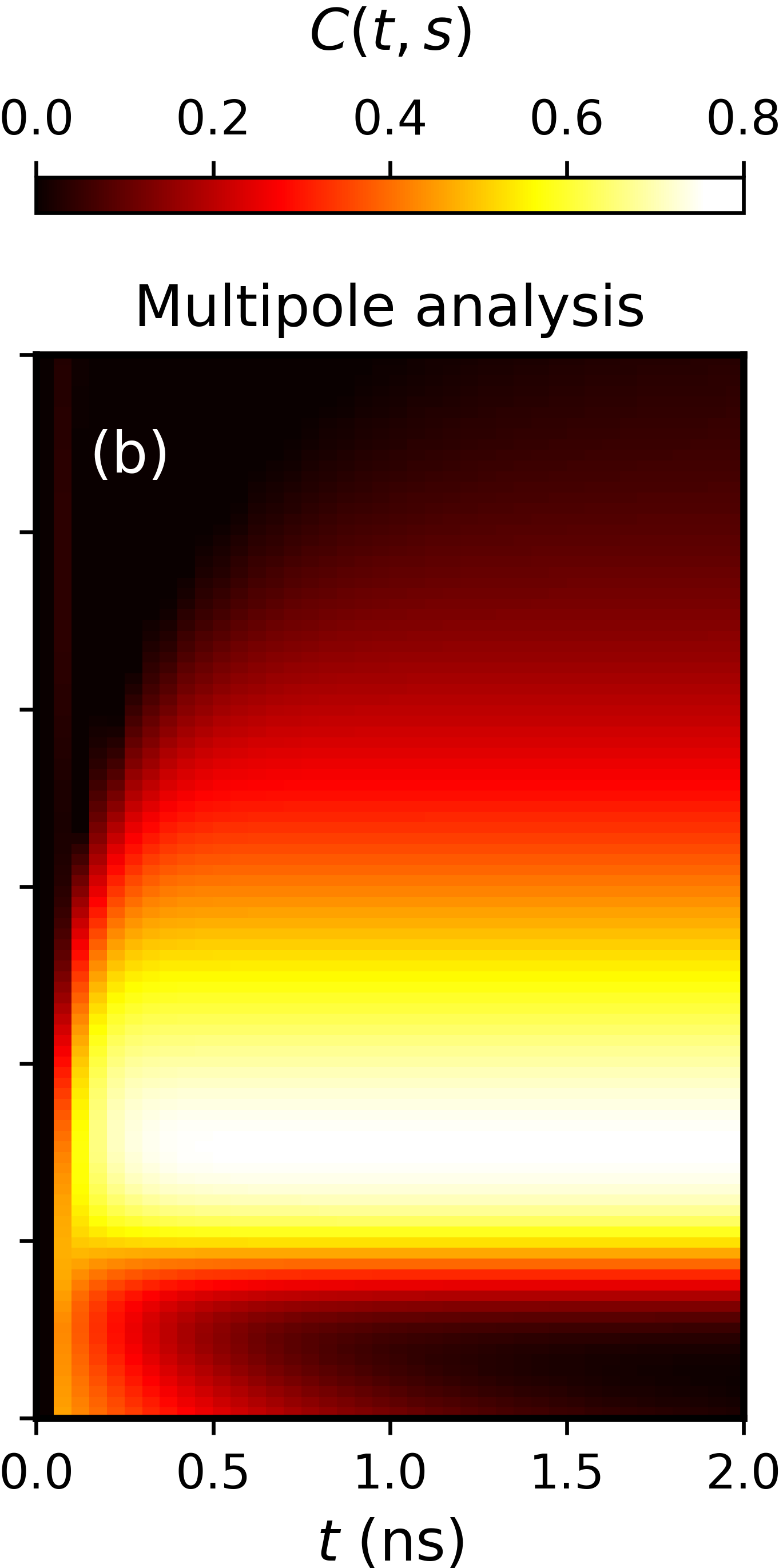}
	\caption{(Color online) Comparison of the dipole and multipole analyses of the transient concurrence of the two-qubit system for qubits prepared in the initial state $\rho_{0} = |3\rangle\langle 3|$ for different values of MNP-QD surface-to-surface distance $s$, and a AgNP of radius $r = 30$ nm.
	}\label{f4}
\end{figure}

\begin{figure}
	\centering 
	\includegraphics[width = 0.35\textwidth, height=0.30\textheight]{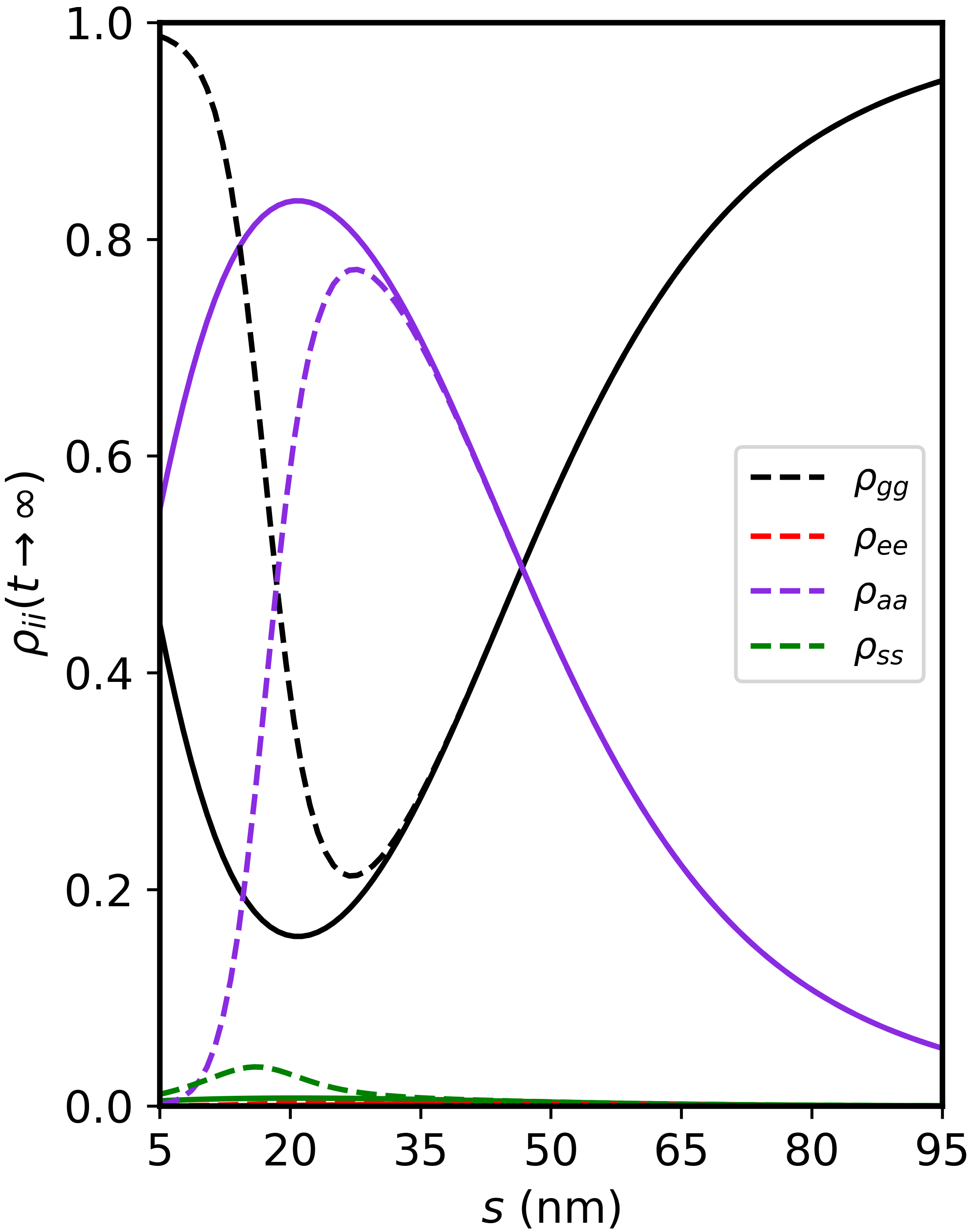}
	\caption{(Color online) Dependence of the steady-state populations, $\rho_{ii}(t\rightarrow\infty), i = g, e, a, s,$ of the effective two-qubit system on the surface-to-surface distance $s$ between each QD qubit and the MNP. Solid lines: Dipole analysis. Dashed lines: Multipole analysis.
	}\label{f8}
\end{figure}

\begin{figure}
	\centering 
	\includegraphics[width = 0.35\textwidth, height=0.30\textheight]{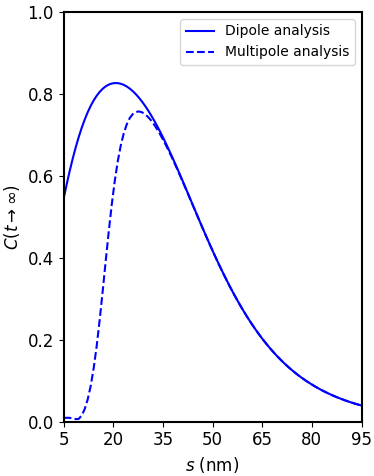}
	\caption{(Color online) Dependence of the dipole and multipole analyses of the stationary concurrence on the MNP-QD surface-to-surface distance $s$, for a AgNP of radius $30$ nm. 
	}\label{f5}
\end{figure}

Figs.~\ref{f3} (a) and (b), as well as Figs.~\ref{f4} (a) and (b) show a side-by-side comparison of the contributions of dipole and multipole plasmons to the generation of entanglement in the coupled qubits for representative values of $s$ in the range: $s = 5$ nm to $s = 95$ nm. 
In Fig.~\ref{f3}, the qubits are prepared in the ground state $|1\rangle$, corresponding to the initial density matrix $\rho_{0} = |1\rangle\langle 1|$. This state then evolves and attains a partially entangled state, with the density matrix $\rho(t\rightarrow\infty) \approx \rho_{gg}|1\rangle\langle 1| + \rho_{aa}|a\rangle\langle a| + \rho_{ss}|s\rangle\langle s|$, with $\rho_{ee}(t\rightarrow\infty) \approx 0$, as shown in Fig.~\ref{f8}.
This results from upward transitions from the ground state to the bi-excited state via the symmetric excitation rate $\Omega_{s}$, and downward transition to the symmetric state via $\Omega_{s}$, with the superradiant rate $\gamma_{s}$, as shown in Fig.~\ref{f2}. However, since $\gamma_{s}>>\gamma_{a}$, the population of the entangled state can be trapped in the antisymmetric state via the detuning rate $\Delta_{-}/2$, which couples $|s\rangle$ to $|a\rangle$, as shown in Fig.~\ref{f2}. 
Thus, the degree of entanglement shown in Fig.~\ref{f5} is due to the antisymmetric state population shown in Fig.~\ref{f8}. Fig.~\ref{f8} also shows that the dipole analysis over-predicts $\rho_{aa}(t\rightarrow\infty)$ and under-predicts $\rho_{gg}(t\rightarrow\infty)$ and $\rho_{ss}(t\rightarrow\infty)$ in the regime $s<r$ where multipolar contributions are dominant. As shown in Fig.~\ref{f8}, as $s$ increases, $\rho_{gg}(t\rightarrow\infty)$ decreases until near $s=r$ where $\rho_{gg}(t\rightarrow\infty)$ reaches a minimum value and $\rho_{aa}(t\rightarrow\infty)$ attains a maximum value. However, at $s>r$, $\rho_{gg}(t\rightarrow\infty)$ begins to increase while $\rho_{aa}(t\rightarrow\infty)$ decreases, due to weakening of the mediated interactions.

The analyses in Fig.~\ref{f3} show that when $s < r$, the indirect excitation of dark modes leads to suppression of the degree of entanglement (red and dark regions in Fig.~\ref{f3} (a) versus those in Fig.~\ref{f3} (b) at $s < r$). The decrease in concurrence in this regime is due to an increase in the population of the ground state, $|g\rangle$ (see Fig. \ref{figA1}: (a) and (b)), as a result of a higher contribution of multipolar modes in the mediated interactions. 
However, as $s\rightarrow r$ (see Fig. \ref{figB1}: (a) and (b)), dipolar modes begin to dominate the plasmon-mediated interactions of the two-qubit dynamics, and the population of the ground state decreases. This leads to an increase in the concurrence due to an increase in the population of the anti-symmetric state, $|a\rangle$. Although the concurrence is still limited by the presence of dark modes (yellow and white regions in Fig.~\ref{f3} (a) versus those in Fig.~\ref{f3} (b) near $s = r$), their suppression of entanglement generation is no longer as pronounced. This leads to an optimal concurrence, $0.7 < C(t\rightarrow\infty,s\rightarrow r) < 0.8$. 
On the other hand, when $s>r$, the mediated interactions---$\tilde{\Gamma}_{ij}$ and $\tilde{G}_{ij}$---are mostly dominated by the bright, dipolar mode, leading to a good agreement between the dipole and multipole analyses. A complete agreement between the two analyses is achieved when $s\ge 2r$. However, at these distances, the mediated interactions are weakened, leading to a decrease in the concurrence (red and dark regions in Fig.~\ref{f3} (a) versus those in Fig.~\ref{f3} (b) at $s \ge 2r$). 

In Fig.~\ref{f4}, the qubits are prepared in the state $|3\rangle$, corresponding to the initial density matrix $\rho_{0} = |3\rangle\langle 3|$. 
At $t = 0$, in the regime $s <r$ (Figs.~\ref{f4} (a) and (b)), this state has no degree of entanglement, but it evolves rapidly ($t\sim 0.1$~ns) to an entangled state with $C(t) > 0$ (see also Fig. \ref{figA1}: (c) and (d)). 
In the dipole analysis, Fig.~\ref{f4} (a), this entangled state evolves slowly to states with higher degrees of entanglement (yellow and white regions in Fig.~\ref{f4} (a) at $s<r$). However, the multipole analysis, Fig.~\ref{f4} (b), reveals the correct trend, which is that the entangled state evolves faster to states with lower degrees of entanglement (red and dark regions in Fig.~\ref{f4} (b) at $s<r$). 
Thus, we refer to the regime $s<r$ as the \emph{strong-coupling regime}, where the degree of entanglement decays due to strong coupling of the qubits to dark modes. 
In the regime $s \rightarrow r$ (Figs.~\ref{f4} (a) and (b)), the initial state also evolves fast to an entangled state with $C(t) < 0.5$, from which it evolves
to states with higher degrees of stationary entanglement only---yellow and white regions in Fig.~\ref{f4} as $s \rightarrow r$ (see also Fig. \ref{figB1}: (c) and (d)). 
We therefore refer to this regime as the \emph{intermediate-coupling regime}, where coupling to dark modes can limit the concurrence (Fig.~\ref{f4} (a) versus Fig.~\ref{f4} (b) as $s \rightarrow r$) but does not lead to entanglement decay, i.e., absence of dark regions. 
Finally, we have the \emph{weak-coupling regime}, $s>r$, where multipolar effects have become negligible but the concurrence begins to decrease due to weak, mediated interactions (red and dark regions in Figs.~\ref{f4} (a) and (b) at $s>r$). 

As shown in Figs.~\ref{f3} and \ref{f4}, the long-time behavior of the transient concurrence is the same regardless of the initial state of the qubits. This is best captured by examining the stationary concurrence, $C(t\rightarrow\infty)$, of the MNP-coupled qubits, as shown in Fig.~\ref{f5}. In the strong-coupling regime, $s < r$, the dipole analysis over-predicts the stationary concurrence, leading to a peak value of $C(t\rightarrow\infty) > 0.8$ around $s = 25$ nm, as shown in Fig.~\ref{f5}. 
However, due to the presence of dark modes in this regime, the multipole analysis in Fig. \ref{f5} shows that $C(t\rightarrow\infty)$ is highly suppressed, reaching a peak value of $C(t\rightarrow\infty) < 0.8$ near $s = 25$ nm. 

As $s\rightarrow r$, the contribution from dark modes fades and
the QD-MNP-QD system transitions from strong to intermediate coupling, where $C(t\rightarrow\infty)$ reaches an optimal value around $s =r$ in the multipole analysis in Fig.~\ref{f5}. 
When $s > r$, the system transitions to the weak-coupling regime, where both the dipole and multipole analyses agree, as shown in Fig.~\ref{f5}, owing to dominant contributions from dipolar plasmons, but $C(t\rightarrow\infty) \rightarrow 0$ due to weak mediated interactions. 

\subsection{Varying MNP size}\label{subsection4.2}
	We now consider the limit $s=r$ only and compare the dipole and multipole analyses of the concurrence as $r$ increases from $r = s = 5$ nm to $r = s = 95$ nm. The qubits are initialized in the ground state, $\rho_{0} = |1\rangle\langle 1|$.
 \begin{figure}
	\centering 
	\includegraphics[width = 0.22\textwidth]{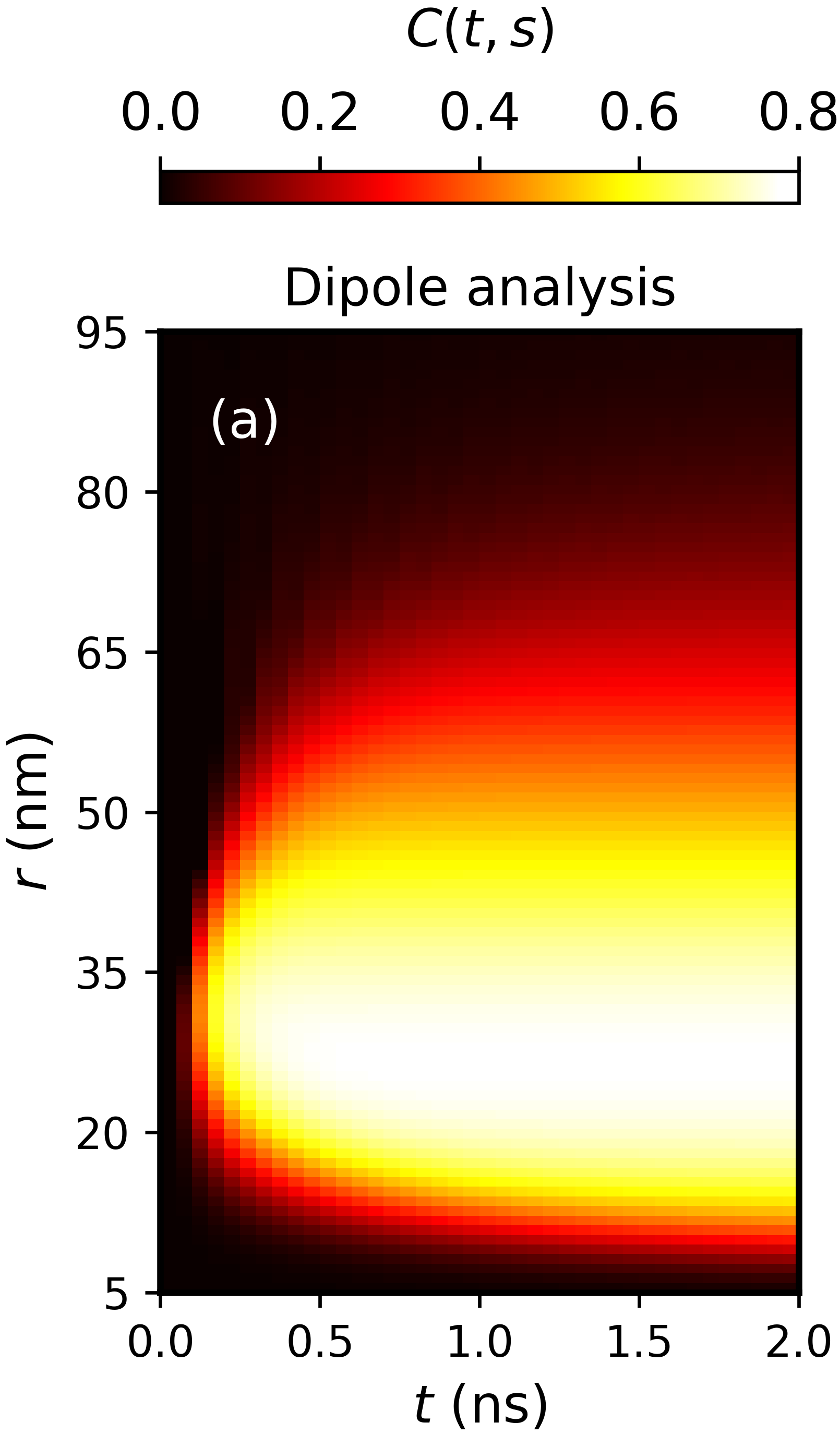}~~~~~~
    \includegraphics[width = 0.18\textwidth, height=0.284\textheight]{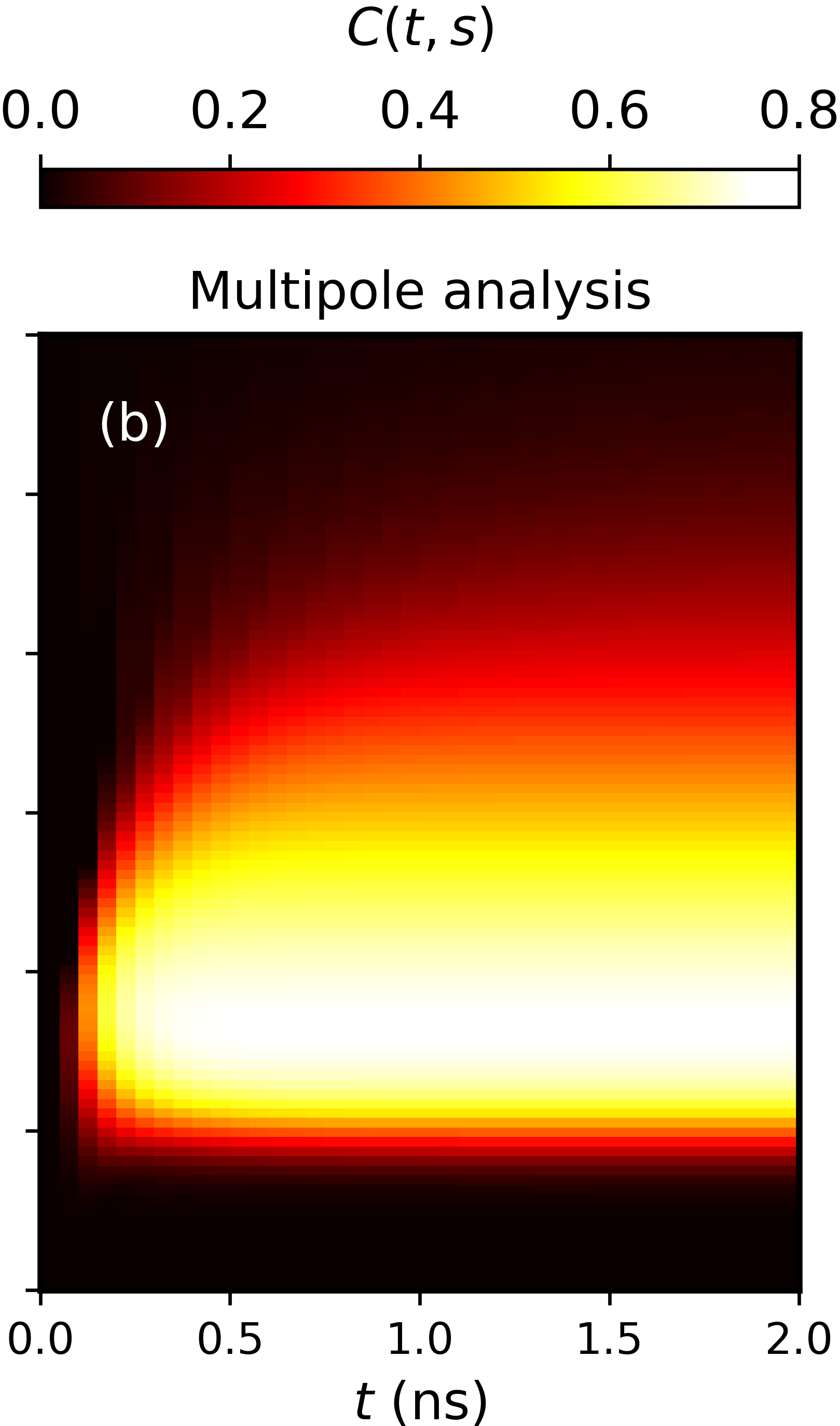}
	\caption{(Color online) Size dependence of the transient concurrence of the two-qubit system for qubits prepared in the initial state $\rho_{0} = |1\rangle\langle 1|$ for the (a) dipole analysis, and (b) multipole analysis, in the limit $s = r$. 
	}\label{f6}
\end{figure}

As shown in Fig.~\ref{f6}, multipolar nonlocal effects lead to size-induced complete suppression of the plasmon-mediated entanglement at $r = s \le 20$ nm (dark regions in 
Fig.~\ref{f6} (a) versus those in Fig.~\ref{f6} (b) at $r \le 20$ nm). 
This regime therefore corresponds to the strong-coupling regime where multipolar modes dominate the mediated interactions. As $r$ increases, for example, Fig.~\ref{f6} (b) in the regime $20$ nm $<r < 50$ nm, nonlocal effects begin to diminish through their dependence on $r$ but dark modes persist due to their dependence on both 
$r$ and $s$. Thus, the degree of entanglement is not suppressed in this regime but it is limited (yellow and white regions in Fig.~\ref{f6} (a) versus those in Fig.~\ref{f6} (b) at $20$ nm $<r < 50$ nm), corresponding to the intermediate-coupling regime. 
Hence, without multipolar corrections, the dipole approximation leads to an over-prediction of the degree of entanglement in these two regimes. 

As $r$ increases beyond $r \approx 50$ nm, the dipole and multipole analyses agree (red and dark regions in Fig.~\ref{f6} (a) versus those in Fig.~\ref{f6} (b) at $r >50$ nm) but the increase in size-dependent plasmon damping together with weak mediated interactions results in a decrease in the concurrence.

\section{Quantum Fisher information}\label{section5}
In this section, we discuss the possible use of plasmon-mediated entangled states in quantum sensing. 
The sensitivity of a quantum state of two qubits, with the density matrix $\rho(\phi)$, to minute changes in the relative phase $\phi$ between the qubits can be determined using the quantum Fisher information (QFI) \cite{Hylus2012}.  
The QFI, denoted as $\mathcal{F}_Q$, bounds the precision in the estimation of $\phi$, via $\Delta\phi \ge 1/\sqrt{m\mathcal{F}_Q}$, for $m$ repeated measurements, according to the Cramer-Rao bound \cite{Hylus2012}. It is given in Refs.~\cite{Hylus2012,Toth2024,Wang2015} as
\begin{equation}
\mathcal{F}_Q = 2 \sum_{i, j} \frac{(\lambda_i - \lambda_j)^2}{\lambda_i + \lambda_j} |\langle \psi_i | H | \psi_j \rangle|^2,
\end{equation}
where $\lambda_i$ and $\lambda_j$ are the eigenvalues of the transformed density matrix $\rho(\phi) = e^{-i\phi H}\rho e^{i\phi H}$ of the probe density matrix $\rho$, $|\psi_i\rangle$ and $|\psi_j\rangle$ are the corresponding eigenvectors, and $H$ is a Hermitian operator, representing a linear operation on the qubits. Here, we consider the generator $H = \frac{1}{2}(\sigma_{z} \otimes I_{2} - I_{2} \otimes \sigma_{z}$), which ensures that the two qubits have opposite phase shifts, i.e., $+\phi/2$ and $-\phi/2$, respectively.
 \begin{figure}
	\centering 
	\includegraphics[width = 0.45\textwidth, height=0.35\textheight]{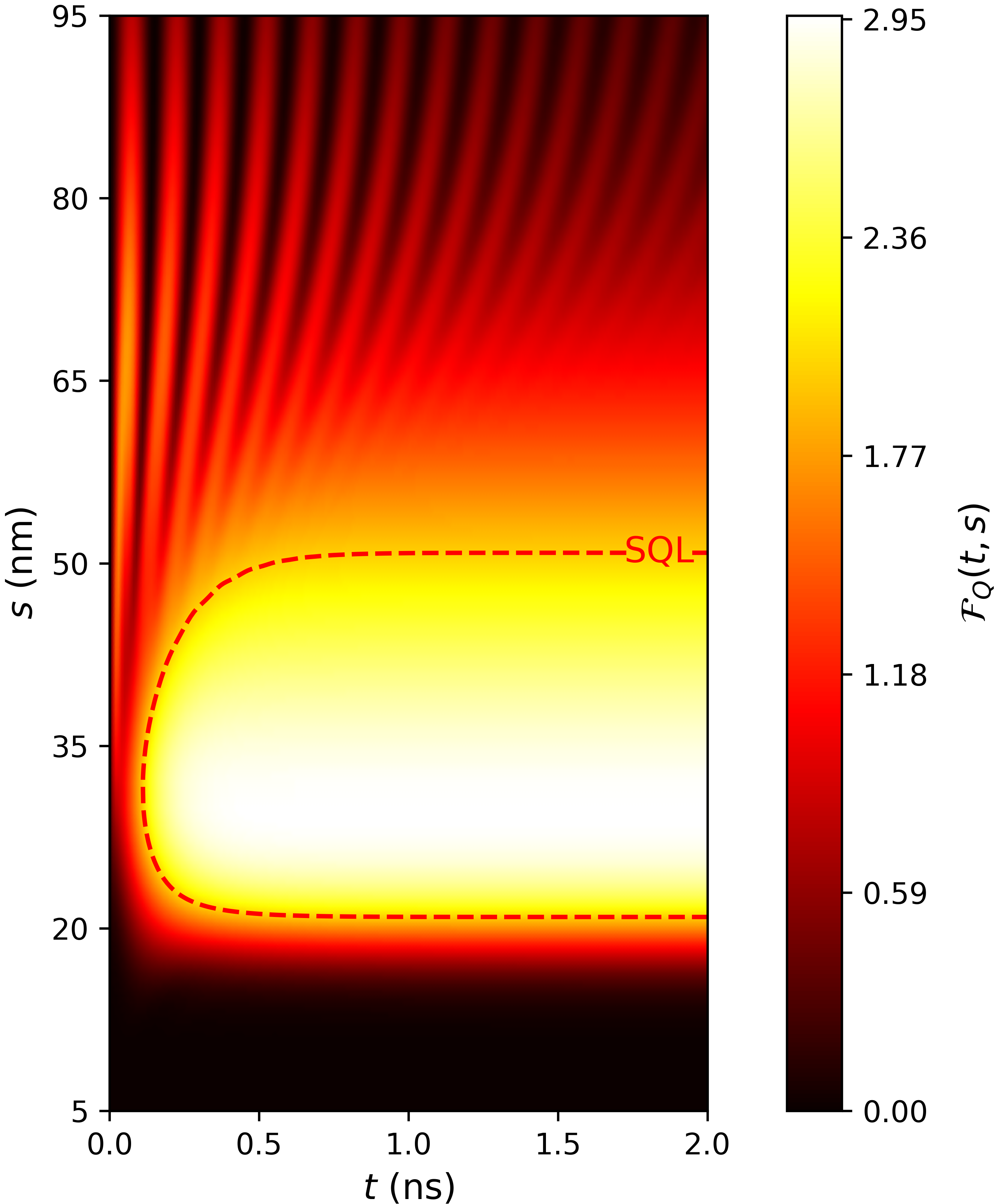}
	\caption{(Color online) Multipole analysis of the QFI in the limit $s=r$ for qubits prepared in the initial state $\rho_{0} = |1\rangle\langle 1|$.
	}\label{f7}
\end{figure}

The dynamics of the QFI in the regime $s=r$ for qubits initially prepared in the ground state is shown in Fig.~\ref{f7}. The Heisenberg limit (HL), with a QFI of $\mathcal{N}^{2}$, with $\mathcal{N}$ being the number of qubits, is the optimal QFI obtainable with a maximally entangled state \cite{Toth2024}. 
In Fig.~\ref{f7}, we do not indicate the HL ($\mathcal{F}_Q = 4$) since it is unattainable with the partially-entangled states we have generated. 
However, in the intermediate coupling regime, between $s = r = 20$ nm and $s = r = 50$ nm, a QFI $\mathcal{F}_Q(t\rightarrow\infty,s)$, greater than the standard quantum limit (SQL)---the QFI of unentangled product states ($\mathcal{F}_Q = 2$)---can be generated using the plasmon-mediated partially-entangled states (regions inside the SQL contour line in 
Fig.~\ref{f7}). 
Fig.~\ref{f7} shows that an optimal QFI, $\mathcal{F}_Q(t\rightarrow\infty,s) \approx 2.95$, is achievable around $r = s= 30$ nm. Thus, mediated entanglement can be optimized to generate partially-entangled states that can outperform classical sensing limits. 

At either $s = r \le 20$ nm (strong-coupling regime) or $s = r \ge 50$ nm (weak-coupling regime). Fig.~\ref{f7} shows that the QFI due to the mediated entanglement at $t\rightarrow\infty$ is below the SQL (regions outside the SQL contour line in Fig. \ref{f7}), since a larger population of mixed states contributes to the QFI in these regimes. In the weak-coupling regime, the QFI oscillates with time due to transient effects that become pronounced as the dark modes die out, leading to the alternating red and black stripes (crests and troughs of the oscillation) in Fig.~\ref{f7} at $s = r \ge 50$ nm. 

\section{Conclusion}\label{section6}
We have extended the plasmonic mediation of entanglement to the nonlocal and multipole regime. We found that dark modes conspire to limit entanglement generation, especially at QD-MNP surface-to-surface distances less than the MNP radius and at certain MNP sizes, highlighting the potential pitfalls of the dipole approximation, such as the over-prediction of the mediated entanglement. 
We identified three coupling regimes: strong coupling, intermediate coupling, and weak coupling. In the strong-coupling regime, multipolar effects must be taken into account in the analysis of the mediated entanglement because of significant contributions from dark modes. In the intermediate coupling regime, multipolar effects begin to diminish, but their presence can still limit entanglement generation, whereas in the weak coupling regime, the dipole and multipole predictions agree, though the degree of entanglement starts to decrease due to weak qubit-qubit-mediated interactions. 
We also found that our study informs a careful selection of the coupling distance and MNP size through which quantum sensing can be carried out to obtain an enhancement in the phase estimation precision. 
This work may help in the design of nanophotonic devices that reliably generate entanglement at the nanoscale for quantum technology applications, such as quantum phase sensing, as we have theoretically demonstrated.
Although we investigated nanosphere mediators, future research directions might employ similar approaches to account for multipolar nonlocal effects in entanglement generation mediated by other nanostructures such as nanorods and nanoshells.

\section*{Acknowledgements}
This research was financially supported by the Department of Science, Technology and Innovation (DSTI) through the South African Quantum Technology Initiative (SA QuTI), Stellenbosch University (SU), the National Research Foundation (NRF), and the Council for Scientific and Industrial Research (CSIR).

\appendix
\section{Population dynamics} 
\setcounter{figure}{0}
\renewcommand{\thefigure}{A\arabic{figure}}

\begin{figure}[h!]
    \centering
    \includegraphics[width=0.38\textwidth]{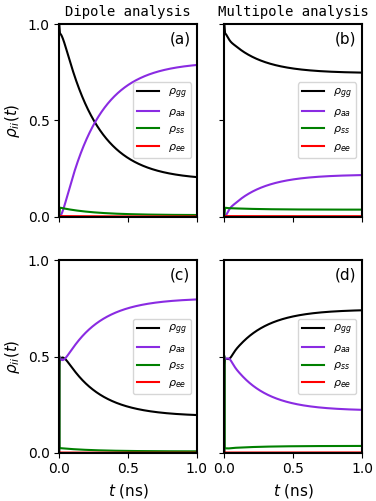}
    \caption{Dynamics of the populations, $\rho_{ii}(t)~(i = g, a, s, e)$, of the MNP-coupled two-QD qubit system at $s = r/2 = 15$ nm. (a) and (b): $\rho_{0} = |1\rangle\langle 1|$. (c) and (d): $\rho_{0} = |3\rangle\langle 3|$}
    \label{figA1}
\end{figure}

\begin{figure}[h!]
    \centering
    \includegraphics[width=0.38\textwidth]{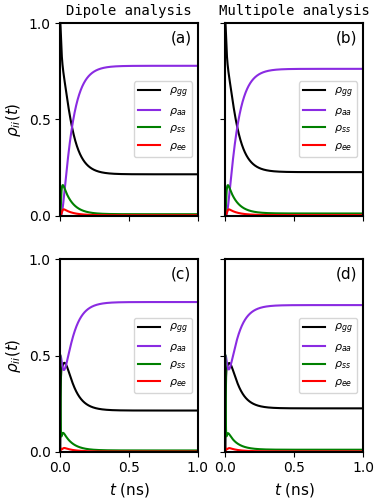}
    \caption{Dynamics of the populations, $\rho_{ii}(t)~(i = g, a, s, e)$, of the MNP-coupled two-QD qubit system at $s = r =30$ nm. (a) and (b): $\rho_{0} = |1\rangle\langle 1|$. (c) and (d): $\rho_{0} = |3\rangle\langle 3|$}
    \label{figB1}
\end{figure}

\newpage 

\bibliographystyle{unsrt}
\bibliography{manuscript.bib}
\end{document}